\documentclass[9pt,conference]{IEEEtran}
\usepackage{amssymb,amsthm,amsmath,array}
\usepackage{graphicx}
\usepackage[caption=false,font=footnotesize]{subfig}
\usepackage{xspace}
\usepackage[sort&compress, numbers]{natbib}
\usepackage{stmaryrd}
\usepackage{xcolor}
\usepackage{mathtools}
\usepackage{float}
\usepackage{textcomp}
\UseRawInputEncoding

\begin{document}
\title{Democratizing AI Governance: Balancing Expertise and Public Participation}
\author{\IEEEauthorblockN{
        Lucile Ter-Minassian \IEEEauthorrefmark{1}
    }
    \IEEEauthorblockA{
        \IEEEauthorrefmark{1} University of Oxford }
}
\maketitle
\begin{abstract}
    The development and deployment of artificial intelligence (AI) systems, with their profound societal impacts, raise critical challenges for governance. Historically, technological innovations have been governed by concentrated expertise with limited public input. However, AI's pervasive influence across domains such as healthcare, employment, and justice necessitates inclusive governance approaches. This article explores the tension between expert-led oversight and democratic participation, analyzing models of participatory and deliberative democracy. Using case studies from France and Brazil, we highlight how inclusive frameworks can bridge the gap between technical complexity and public accountability. Recommendations are provided for integrating these approaches into a balanced governance model tailored to the European Union, emphasizing transparency, diversity, and adaptive regulation to ensure that AI governance reflects societal values while maintaining technical rigor. This analysis underscores the importance of hybrid frameworks that unite expertise and public voice in shaping the future of AI policy.
\end{abstract}

\subsection*{Introduction}
Technological innovation has historically emerged from concentrated centers of expertise, developed by a limited group of specialists with minimal public oversight or legal frameworks (e.g. printing press, nuclear weapons). The development of artificial intelligence—defined here as machine learning models trained on data that can be deployed for real-world applications, encompassing both supervised and unsupervised approaches, whether predictive or generative—follows this pattern, with development concentrated among a small number of private companies and research institutions \cite{moura2024}.
However, AI systems are increasingly deployed in domains that affect fundamental aspects of human life, from healthcare and education to employment and criminal justice \cite{west2018, obermeyer2019, angwin2016}. This broad societal impact, combined with AI's potential to reshape power structures and decision-making processes, raises critical questions about its governance. A stakeholder-based analysis suggests that those affected by AI systems should have a voice in shaping both their development and deployment.
Yet this proposition faces significant challenges. The technical complexity of AI systems creates a substantial knowledge gap between experts and the general public. This disconnect raises fundamental questions about the feasibility of democratic participation in AI governance: How can meaningful public input be incorporated into decisions that require deep technical understanding? What mechanisms could bridge the divide between scientific expertise and democratic legitimacy?
Democratic governance, defined here as decision-making processes that ensure representation, accountability, and participation from affected stakeholders, presents both opportunities and challenges in the context of AI. This analysis will examine competing schools of thought: first exploring arguments for democratic governance, then considering the case for expert-led oversight. Ultimately, we will interrogate the possibility of integrating direct democratic participation into AI governance frameworks. Drawing on case studies from France and Brazil, we propose recommendations for a balanced governance model tailored to European Union countries.

\section{The Case for Democratic AI Governance: A Stakeholder-Based Analysis}
\subsection{Universal Stakeholdership and Access}
The democratization of AI technology represents a significant departure from historical patterns of technological innovation. Unlike previous transformative technologies such as nuclear weapons or early printing presses, which were characterized by centralized control and limited access, AI exhibits unprecedented openness through open-source initiatives.
This openness manifests across three distinct dimensions. First, access to AI applications has become widespread, with tools ranging from chatbots to image generators freely available to any internet user. Second, AI development appears accessible; while anyone with basic computational resources can fine-tune language models using open-source frameworks, the inherent complexity and opacity of AI systems preserve the traditional knowledge barrier between experts and the general public \cite{arrieta2020}. Third, AI's influence extends across critical domains, including healthcare, insurance, and media consumption, creating a unique context where every citizen is both an active and passive stakeholder in AI's deployment \cite{west2018, obermeyer2019, angwin2016}.
This three-tiered reality—universal access, technically-constrained development, and universal impact—creates a unique governance challenge. Despite widespread use and impact, decision-making processes remain concentrated among tech companies and policymakers, with limited public participation. The European Union's AI Act negotiations exemplify this democratic deficit, where civil society, despite representing the interests of affected populations, remains largely excluded from crucial policy formations \cite{veale2021}. This creates a concerning misalignment between AI's universal stakeholdership and its concentrated governance structures.

\subsection{Democratic Legitimacy in an Era of Declining Trust}
Contemporary democratic institutions face a crisis of legitimacy, evidenced by declining public trust across OECD nations. Empirical data indicates a roughly 20\% decrease in confidence in governmental bodies over two decades, with Eurobarometer surveys showing trust in national governments falling below 35\% in numerous EU member states \cite{oecd2024}. Within this context, democratic AI governance becomes crucial for two reasons:

\textbf{1. Legitimacy and Accountability:} Public participation in AI governance ensures policy alignment with collective interests rather than narrow corporate or technical objectives. Arras and Braun \cite{arras2020} experimentally show that inclusive consultation processes can enhance regulatory legitimacy.

\textbf{2. Transparency and Trust:} AI's inherent opacity and pervasive influence necessitate democratic oversight to prevent public skepticism and resistance. O'Neil et al. demonstrate how opaque algorithmic systems can perpetuate inequality when developed without adequate public scrutiny \cite{oneil2016}.

\subsection{Limitations of Expert-Centric Governance}
Expert-only governance models present several critical limitations:

\textbf{1. Representation Deficits:} Exclusive technical governance risks overlooking broader ethical and social implications, potentially creating policies misaligned with public priorities. Zuboff's analysis of early internet governance illustrates how expert-centric approaches failed to anticipate crucial societal challenges like misinformation and privacy concerns \cite{zuboff2019}.

\textbf{2. Interdisciplinary Necessity: }To effectively address the complex societal implications of AI, it is essential to adopt an interdisciplinary approach that bridges technical expertise with ethical and policy considerations. Coeckelbergh's concept of "democracy deficit" highlights how technocratic approaches often neglect the inherently political nature of AI-related decisions. \cite{coeckelbergh2020} Cave et al. emphasize the importance of integrating perspectives from computer science, ethics, and policy to address AI's multifaceted challenges. \cite{cave2019}

\subsection{Economic and Civil Liberty Implications}
Since AI serves as a key driver of economic growth and a powerful tool with societal implications, democratic governance, characterized by its emphasis on inclusivity and collective decision-making, may offer several advantages for addressing economic and civil liberty concerns. First, it may promote economic equity by ensuring a more equitable distribution of AI-driven benefits across society, particularly supporting small and medium-sized enterprises \cite{arras2020}. Second, democratic frameworks may enhance rights protection by enabling public participation, which could help balance security requirements with civil liberties such as privacy and freedom of expression \cite{cheong2024}. Finally, they may mitigate conflicts of interest by reducing the disproportionate influence of entities that primarily profit from AI development and deployment, fostering a more inclusive and balanced governance process.

\section{The Scientific Imperative: Why AI Governance Requires Expert Leadership}

\subsection{Scientific Method and Democratic Process: Fundamental Incompatibilities}
Klein's analysis in Le goût du vrai presents a fundamental critique of democratic approaches to scientific governance. \cite{klein2020} He articulates three critical tensions between scientific advancement and democratic processes:

\textbf{1. Epistemological Foundation:} Scientific truth emerges through empirical validation and rigorous methodology, not through consensus or popular agreement. This epistemological framework stands in direct opposition to democratic principles of collective decision-making.

\textbf{2. Expertise Hierarchy:} While democratic systems presuppose equality of voice, scientific progress necessitates the privileging of expert knowledge. This hierarchical structure of expertise, as elaborated by Collins and Evans in Rethinking Expertise \cite{collins2007}, is essential for maintaining scientific rigor and preventing the dangerous flattening of specialized knowledge.

\textbf{3. Methodological Misalignment}: Public interpretation of scientific processes often fails to grasp the necessity of uncertainty and debate within scientific methodology, potentially undermining crucial aspects of scientific advancement.

The COVID-19 pandemic response provides a compelling empirical example: epidemiological expertise necessarily superseded public opinion in implementing lockdown measures, demonstrating how scientific imperatives can override democratic processes in crisis situations.

\subsection{Misinformation and Public Discourse}

The proliferation of AI misinformation in public discourse presents a significant challenge to democratic governance models. Cave's analysis in Bridging near- and long-term concerns about AI \cite{cave2019} identifies two primary distortions that shape public understanding. On one hand, exaggerated threat narratives amplify potential risks beyond their empirical basis, fostering undue fear and misunderstanding. Conversely, popular media often promotes unrealistic optimism, presenting oversimplified or overly positive views of AI capabilities. Together, these distortions hinder informed decision-making and complicate efforts to address the societal implications of AI through democratic means.

These distortions in public understanding can lead to policy decisions that fail to address actual technological challenges and risks. Furthermore, due to their inclusive nature, democratic processes may also be susceptible to populist pressures, potentially compromising essential strategic and ethical considerations in AI development.

\subsection{Temporal Constraints and Decision-Making Efficiency}

The rapid pace of AI development creates temporal pressures that conflict with democratic deliberation. The speed of technological advancement consistently outpaces democratic decision-making processes, leading to a persistent innovation-governance gap where regulatory responses lag behind new developments. Additionally, high-stakes decisions in AI development often demand immediate, expert-informed responses, which democratic processes are not always equipped to efficiently facilitate. These temporal and procedural constraints suggest that expert-led governance frameworks may be better suited to addressing the immediate challenges of AI development and deployment.

\section{Towards Balanced AI Governance: Integrating Democratic Participation with Expert Knowledge}
While technical expertise remains crucial, integrating democratic participation in AI governance can help ensure that policy decisions are aligned with societal values, fostering legitimacy and public trust in the outcomes. In this section, we explore two key models of democratic initiatives: participatory and deliberative democracy. Through the examination of two case studies—the French Citizens’ Convention on Climate (CCC) and the Brazilian AI Framework—we analyze how these approaches operate in practice. We conclude by drawing lessons from these examples and applying them to the context of AI governance, emphasizing how democratic engagement can enhance inclusivity, transparency, and accountability in addressing complex, technical challenges.
\\
\\
In the remainder of the paper, we will use the notions of participatory democracy and deliberative democracy. Participatory democracy emphasizes broad public engagement and direct decision-making, often through methods like voting, surveys, or public comments, aiming for maximum inclusion and citizen power. In contrast, deliberative democracy focuses on informed debate and consensus-building through structured dialogue, typically in smaller groups facilitated by trained moderators. While participatory democracy prioritizes the breadth of participation, deliberative democracy emphasizes depth, requiring more time and interaction with technical experts to achieve well-reasoned, collective decisions. These approaches differ in their goals: participatory democracy seeks widespread involvement, while deliberative democracy aims for thoughtful, consensus-driven outcomes.

\subsection{Accessibility of AI Governance}

The governance of AI systems often presents fewer technical barriers than commonly assumed. While some aspects undeniably require deep technical expertise, many critical governance decisions focus on universally relatable societal impacts. Dryzek \cite{dryzek2019} advocates for participatory and decentralized governance, combining scientific expertise with public values and social considerations to address complex challenges, such as those posed by environmental issues. Similarly, Pasternak et al. \cite{pasternak2024} demonstrate how teaching core principles of complex scientific topics can empower policymakers to make informed decisions. They detail their experience in designing and delivering courses aimed at equipping policymakers to understand and apply scientific evidence effectively. Drawing from these perspectives, we argue that effective AI governance demands not only an understanding of core technical concepts, such as AI training methodologies, robustness, and explainability principles, but also a deep awareness of their societal implications for privacy, labor markets, and ethical considerations.

\subsection{Democratic Models in Practice: the French Citizens' Convention on Climate (CCC)}

Participatory democracy emphasizes broad public consultation and direct decision-making mechanisms, while deliberative democracy prioritizes informed debate through structured expert-citizen dialogue. The historical precedent of the Athenian Boule from 508-507 BCE provides an instructive model: a council of 500 citizens, chosen by lot annually, engaged with complex technical and administrative matters while consulting experts. The Boule met daily in specialized committees, demonstrating how ordinary citizens could effectively participate in technical governance when provided with appropriate structural support.
\\
The French Citizens' Convention on Climate of 2019 \cite{pech2019, eymard2022} (CCC) exemplifies modern application of these principles. The CCC was a democratic experiment that aimed to give citizens a direct voice in shaping climate policy. It brought together 150 citizens selected through sortition to represent the diversity of French society across six demographic dimensions: gender, age, socio-economic background, education level, location type, and province. Participants were tasked with defining measures to reduce greenhouse gas emissions by at least 40\% by 2030, compared to 1990 levels, while considering social justice \cite{pech2019}. 
\\
\\
\textbf{Process and Methodology}
\begin{itemize}
    \item \textbf{Thematic Working Groups: }The assembly was divided into five work groups covering transport, food, consumption, work and production, and housing.
    \item \textbf{Expert Support:} The convention was assisted by a governance committee, legal and technical experts, and professionals in citizen participation and deliberation.
    \item \textbf{Deliberative Approach:} Citizens learned about climate issues, debated, and prepared draft laws over seven sessions spanning several months. 
    \item \textbf{Public engagement through online contributions:} Citizens could submit suggestions through an online platform. Over 1,000 contributions were synthesized and provided to the convention participants, though time constraints limited their thorough consideration. \cite{gaborit2022,knoca2025}
\end{itemize}

All participants received compensation as well as loss of professional income when relevant. Through their deliberations, participants produced 149 distinct policy proposals, showcasing citizens' capacity to meaningfully engage with complex issues and generate concrete solutions. The initiative gained particular significance when the President committed to submitting these proposals "without a filter" through various democratic channels, including referendum, parliamentary vote, or direct implementation. However, ultimately most proposals weren’t implemented in their original form. The government introduced the "Climate and Resilience Law" in 2021, which drew from the CCC’s recommendations but watered down several key proposals. \cite{knoca2025} Beyond its immediate policy implications, the Convention fostered increased public awareness and engagement with climate issues, potentially catalyzing greater societal acceptance and support for climate action measures.

\textbf{Challenges of the CCC} 

The Citizens' Convention on Climate faced challenges balancing deliberative independence and external engagement, as explained by Gaborit et al. \cite{gaborit2022} Internally, tensions arose over expert influence, as participants sometimes perceived experts as steering discussions, particularly when their advice conflicted with citizen priorities. Externally, public contributions were synthesized but only partially utilized due to time constraints. Finally, activist groups lobbied participants near the convention venue, raising overlooked issues but risking undue influence on deliberations. Media interactions amplified public awareness but placed additional pressure on participants to act as both deliberators and public representatives, complicating the focus on internal decision-making. These dynamics highlight the difficulty of maintaining a clear boundary between the assembly's internal deliberations and its broader societal and institutional interactions. \cite{gaborit2022}
\\
\subsection{Democratic Models in Practice: the Brazilian AI Framework}
The Brazilian AI Framework \cite{zanatta2024, digitalpolicy2024, dataprivacy2025} aims to establish a comprehensive regulatory structure for the development, implementation, and use of artificial intelligence systems in Brazil, focusing on protecting fundamental rights, promoting responsible innovation, and ensuring safe and reliable AI systems for the benefit of society. 
\\
The framework was the result of a lengthy development, characterized by extensive deliberation, public engagement, and collaboration. Initially, it consolidated three separate bills (5.051/2019, 21/2020, and 872/2021). In March 2022, a commission was established to draft a new unified law, working over 240 days through meetings, seminars, and public hearings.
\\
The Senate Temporary Committee on Artificial Intelligence (CTIA) played a key role, incorporating 85 amendments into the bill after extensive debates and multiple deadline extensions. Public engagement was central, involving hearings and contributions from stakeholders, such as recommendations from the Centre for Information Policy Leadership (CIPL). 
\\

Key features of the framework include:
\begin{itemize}
    \item \textbf{Mandatory algorithmic impact assessments} for high-risk AI systems developed by the companies. This ensures continuous evaluation of risks and benefits to fundamental rights. \cite{dataprivacy2025}
    \item \textbf{Public consultations for public sector AI:} The Brazilian National Data Protection Authority (ANPD) initiated a public consultation to collect a broad range of contributions for shaping AI guidelines and regulations. This effort aimed to incorporate diverse perspectives from both technical experts and civil society, reflecting the democratic aspirations of the framework. While the consultation highlighted openness, the details of how contributions were synthesized remain limited. \cite{digitalpolicy2024}
    \item \textbf{Public consultations for public sector AI:} The Brazilian National Data Protection Authority (ANPD) initiated a public consultation to collect a broad range of contributions for shaping AI guidelines and regulations. This effort aimed to incorporate diverse perspectives from both technical experts and civil society, reflecting the democratic aspirations of the framework. While the consultation highlighted openness, the details of how contributions were synthesized remain limited. \cite{digitalpolicy2024}
    \item \textbf{Regular dialogue between citizens and authorities:} The creation of the National System for Artificial Intelligence Regulation and Governance (SIA) provides a structural mechanism for ongoing engagement. This system ensures that regulatory approaches to AI remain adaptive and aligned with societal needs, fostering trust and accountability. It signifies a commitment to maintaining citizen involvement as AI technologies evolve. \cite{digitalpolicy2024}
    \item \textbf{Extensive public consultation during development:} The process included public hearings, seminars, and stakeholder input. However, the depth and impact of these consultations on the final legislation are not fully transparent, raising questions about whether public feedback directly influenced the bill's provisions.
\end{itemize}

The bill ultimately passed with widespread support on December 10, 2024. \cite{digitalpolicy2024}

The Brazilian AI Framework stands as a landmark in integrating citizen participation into AI governance, reflecting a balance between stakeholder input and legislative action. However, despite its significant public engagement, the absence of a citizen panel limited its scope as a fully participatory democratic initiative.

\subsection{Lessons for AI governance}

Lessons from the French Citizens' Convention on Climate (CCC) and the Brazilian AI Framework provide valuable insights for creating inclusive, effective AI governance. They show how structured democratic engagement can address complex, technical challenges while also exposing the inherent limits of such approaches.

\paragraph{Inclusive Representation and Diversity}
Ensuring inclusive representation is crucial. The CCC selected participants by sortition to reflect France’s diversity in terms of gender, age, socio-economic background, and geography, ensuring varied perspectives in policy recommendations—an approach AI governance should emulate to capture diverse societal concerns. The Brazilian AI Framework, though it involved public consultation, lacked a mechanism for direct citizen representation. This underscores the importance of actively involving citizens beyond public hearings to empower their voices.

\paragraph{Balancing Expertise and Public Engagement}
Effective governance must bridge knowledge gaps between experts and citizens. The CCC used thematic working groups and expert support so citizens could engage with complex issues like climate policy. Likewise, AI governance can benefit from structured deliberation, allowing citizens and AI experts to collaboratively address topics such as bias, transparency, and accountability. The Brazilian AI Framework faced time constraints and ambiguous feedback channels, highlighting the need for processes that collect and visibly integrate public input.

\paragraph{Challenges of Deliberative Processes}
The CCC encountered difficulties balancing independence and external engagement, including perceived expert influence, lobbying, and media pressure. These challenges point to the need for boundaries and safeguards to protect citizen panels’ integrity. In contrast, the Brazilian AI Framework lacked a dedicated citizen panel, reducing such tensions but also limiting the depth of deliberation. Its reliance on public consultations and expert input made it more expert-driven, illustrating trade-offs between efficiency and democratic participation.

\paragraph{Transparency, Accountability, and Iterative Processes}
Transparency and accountability are vital for building trust in governance. The CCC showed this through live-streamed sessions and published documentation, and AI governance should follow suit by providing public access to deliberations and clear accountability measures. The Brazilian AI Framework’s National System for Artificial Intelligence Regulation and Governance (SIA) facilitates ongoing dialogue between citizens and authorities, supporting iterative processes, continuous input, and periodic reviews of governance strategies as societal needs evolve.

\paragraph{Toward a Nuanced Hybrid Framework}
These lessons suggest that democratic participation in AI governance is both possible and beneficial when structured carefully. The key is designing hybrid frameworks that blend expert knowledge with meaningful public input while managing tensions between technical complexity and democratic engagement. 

Challenges from both the CCC and the Brazilian AI Framework show the need for intentional design to manage external pressures, incorporate public feedback, and balance deliberative independence with transparency. The next section offers specific recommendations for such hybrid frameworks in the European Union, building on existing structures to foster inclusive, transparent, and technically credible AI governance.

\section{Framework Recommendations: Implementing Democratic AI Governance in the European Union}

In this section, we provide recommendations for a comprehensive AI governance framework tailored for EU member states, aligning with the broader provisions of the EU AI Act. While not exhaustive, these recommendations aim to guide the development of governance models that integrate deliberative and participatory elements. The proposed framework is structured around key pillars: representative selection, adaptive governance mechanisms, and sustained public engagement. Together, these pillars offer a pathway for fostering inclusive, transparent, and effective governance of AI systems.

\subsection{Aligning Democratic Governance Models with EU AI Initiatives}

Participatory and deliberative approaches to AI governance at the country level within the EU may provide a robust mechanism for aligning national efforts with overarching EU frameworks, including the measures proposed in the First Draft of the General-Purpose AI Code of Practice \cite{eu2024}. These approaches may reinforce the principles of transparency, accountability, and inclusivity while ensuring the public's voice informs the governance of AI systems.
The First Draft of the General-Purpose AI Code of Practice \cite{eu2024} explicitly highlights the importance of involving civil society, academia, and other non-expert groups to address systemic risks posed by general-purpose AI. For instance, Measure 10.1 calls for the integration of participatory methods in evidence collection. EU member states could explore national initiatives such as citizen assemblies or deliberative panels to contribute to this evidence base, potentially offering diverse perspectives on AI risks and societal impacts.
Additionally, the Code’s emphasis on transparent documentation (Measures 1 and 2) may provide opportunities for participatory approaches to flourish. National governments might consider creating publicly accessible platforms for sharing technical documentation, including risk assessments and acceptable use policies (AUPs), as mandated by the EU AI Act. These platforms could include mechanisms for public feedback and input, allowing citizens to participate directly in the scrutiny and refinement of AI policies.
Incorporating public oversight measures outlined in the Code, such as serious incident reporting (Measure 18) and whistleblowing protections (Measure 19), could align with deliberative practices. Member states might consider establishing independent oversight bodies with citizen representation to review and respond to reported incidents. Similarly, strengthening protections for whistleblowers might enable individuals to safely highlight risks or ethical concerns, fostering trust in the governance process.
Finally, the post-deployment monitoring framework (Measure 11) proposed in the Code \cite{eu2024} could encourage sustained public engagement. Member states might implement participatory mechanisms for ongoing monitoring, such as community-led reviews or participatory audits of AI systems, to ensure that governance efforts remain adaptive to societal needs and technological evolution.
In conclusion, participatory and deliberative initiatives at the national level may not only align with the EU AI Act and the General-Purpose AI Code of Practice but also have the potential to enhance the inclusivity and effectiveness of AI governance. These approaches may help ensure that governance models reflect the diverse perspectives of EU citizens while adhering to the principles enshrined in the EU's broader AI regulatory framework.

\subsection{Temporal Considerations and Adaptive Governance}

A stepwise approach may reconcile rapid AI innovation with democratic deliberation. Drawing on Cowan’s \cite{cowan1997} analysis of technological governance, the framework could initially rely on stronger expert oversight, then gradually integrate broader public input as societal understanding deepens. This transition might coincide with milestones like consensus on model explainability or baseline ethical standards.
To address AI’s rapid evolution potentially outpacing traditional policy-making, adaptive mechanisms are crucial. Iterative regulations, mixed expert-public advisory panels, and agile “sandboxes” allow incremental policy testing and refinement, aligning with technological advances and societal needs. Continuous public engagement can reflect shifting societal values and emerging concerns. Involving diverse stakeholders in ongoing discussions helps policies evolve alongside AI’s progress while staying aligned with public interests, bridging the gap between rapid technological change and slower policy processes.
In summary, adaptive and temporal considerations are pivotal for a governance framework that evolves over time. Gradual implementation, responsive regulations, and iterative engagement can help manage AI’s rapid development while preserving democratic principles.

\subsection*{Conclusion}
This analysis explores the tension between democratic participation and expert governance in AI development. While scientific rigor demands specialized expertise, AI’s deep societal impact calls for public engagement. Real-world examples show that technical complexity does not have to exclude meaningful public input; rather than choosing between expertise and democracy, effective AI governance hinges on their careful integration.
The proposed EU framework demonstrates how structured deliberation and phased public participation can address time-sensitive AI challenges without compromising democratic legitimacy. By uniting random selection, expert guidance, and open engagement channels, it balances the urgency of AI innovation with the need for public trust. As AI continues to transform society, governance must preserve both technical excellence and collective values. Looking ahead, pilot initiatives and ongoing feedback loops can refine these democratic mechanisms, ensuring that AI governance remains inclusive, transparent, and adaptive. The future of AI governance lies in synthesizing expertise and public voice—an approach that strengthens accountability and positions AI as a truly shared endeavor.

\bibliographystyle{IEEEtran}
\bibliography{main}

\begin{thebibliography}{10}
\providecommand{\url}[1]{#1}
\csname url@samestyle\endcsname
\providecommand{\newblock}{\relax}
\providecommand{\bibinfo}[2]{#2}
\providecommand{\BIBentrySTDinterwordspacing}{\spaceskip=0pt\relax}
\providecommand{\BIBentryALTinterwordstretchfactor}{4}
\providecommand{\BIBentryALTinterwordspacing}{\spaceskip=\fontdimen2\font plus
\BIBentryALTinterwordstretchfactor\fontdimen3\font minus \fontdimen4\font\relax}
\providecommand{\BIBforeignlanguage}[2]{{%
\expandafter\ifx\csname l@#1\endcsname\relax
\typeout{** WARNING: IEEEtran.bst: No hyphenation pattern has been}%
\typeout{** loaded for the language `#1'. Using the pattern for}%
\typeout{** the default language instead.}%
\else
\language=\csname l@#1\endcsname
\fi
#2}}
\providecommand{\BIBdecl}{\relax}
\BIBdecl

\bibitem{moura2024}
P.~Jácome~de Moura~Jr., C.~D. dos Santos~Junior, C.~G. Porto-Bellini, and J.~J. Lima Dias~Junior, ``The over-concentration of innovation and firm-specific knowledge in the artificial intelligence industry,'' \emph{Journal of the Knowledge Economy}, 2024, advance online publication.

\bibitem{west2018}
D.~M. West and J.~R. Allen, \emph{How Artificial Intelligence Is Transforming the World}.\hskip 1em plus 0.5em minus 0.4em\relax Brookings Institution, 2018.

\bibitem{obermeyer2019}
Z.~Obermeyer, B.~Powers, C.~Vogeli, and S.~Mullainathan, ``Dissecting racial bias in an algorithm used to manage the health of populations,'' \emph{Science}, 2019.

\bibitem{angwin2016}
J.~Angwin, J.~Larson, S.~Mattu, and L.~Kirchner, ``Machine bias: There's software used across the country to predict future criminals,'' \url{https://www.propublica.org/}, 2016.

\bibitem{arrieta2020}
A.~B. Arrieta, N.~Díaz-Rodríguez, J.~Del~Ser, A.~Bennetot, S.~Tabik, A.~Barbado, and F.~Herrera, ``Explainable artificial intelligence (xai): Concepts, taxonomies, opportunities and challenges toward responsible ai,'' \emph{Information Fusion}, vol.~58, pp. 82--115, 2020.

\bibitem{veale2021}
M.~Veale and F.~Z. Borgesius, ``Demystifying the draft eu artificial intelligence act—analysing the good, the bad, and the unclear elements of the proposed approach,'' \emph{Computer Law Review International}, vol.~22, no.~4, pp. 97--112, 2021.

\bibitem{oecd2024}
OECD, \emph{OECD Survey on Drivers of Trust in Public Institutions – 2024 Results: Building Trust in a Complex Policy Environment}.\hskip 1em plus 0.5em minus 0.4em\relax Paris: OECD Publishing, 2024.

\bibitem{arras2020}
S.~Arras and D.~Braun, ``Stakeholder consultations and the legitimacy of regulatory decision-making: A survey experiment,'' \emph{Regulation \& Governance}, vol.~14, no.~3, pp. 435--456, 2020.

\bibitem{oneil2016}
C.~O'Neil, \emph{Weapons of Math Destruction: How Big Data Increases Inequality and Threatens Democracy}.\hskip 1em plus 0.5em minus 0.4em\relax Crown Publishing Group, 2016.

\bibitem{zuboff2019}
S.~Zuboff, \emph{The Age of Surveillance Capitalism: The Fight for a Human Future at the New Frontier of Power}.\hskip 1em plus 0.5em minus 0.4em\relax PublicAffairs, 2019.

\bibitem{coeckelbergh2020}
M.~Coeckelbergh, \emph{AI Ethics}.\hskip 1em plus 0.5em minus 0.4em\relax MIT Press, 2020.

\bibitem{cave2019}
S.~Cave, J.~Whittlestone, R.~Nyrup, S.~Ó~hÉigeartaigh, and R.~A. Calvo, ``Bridging near- and long-term concerns about ai,'' \emph{Nature Machine Intelligence}, vol.~1, no.~1, pp. 5--8, 2019.

\bibitem{cheong2024}
I.~Y. e.~a. Cheong, ``Particip-ai: A democratic surveying framework for anticipating ai risks and benefits,'' 2024.

\bibitem{klein2020}
Ã.~Klein, \emph{Le g\^out du vrai}.\hskip 1em plus 0.5em minus 0.4em\relax Éditions Gallimard, 2020.

\bibitem{collins2007}
H.~Collins and R.~Evans, \emph{Rethinking Expertise}.\hskip 1em plus 0.5em minus 0.4em\relax University of Chicago Press, 2007.

\bibitem{dryzek2019}
J.~S. Dryzek, \emph{The Politics of the Earth: Environmental Discourses}, 4th~ed.\hskip 1em plus 0.5em minus 0.4em\relax Oxford University Press, 2019.

\bibitem{pasternak2024}
N.~Pasternak, P.~Almeida, R.~J. Seixas, and E.~M. Fonseca, ``Teaching scientific evidence and critical thinking for policy making,'' \emph{PLoS ONE}, vol.~19, no.~4, p. e0293819, 2024.

\bibitem{pech2019}
T.~Pech and L.~Tubiana, ``Citizens' convention on climate: Governance and procedure,'' 2019.

\bibitem{eymard2022}
L.~Eymard, A.~Fabre, and S.~Strauss, ``Lessons from the french citizens' convention for climate,'' \emph{HAL ENPC}, 2022, hal-03119539v29.

\bibitem{gaborit2022}
M.~Gaborit, L.~Jeanpierre, and R.~Rozencwajg, ``Les frontières négociées des assemblées citoyennes. le cas de la convention citoyenne pour le climat (2019-2020),'' \emph{Participations}, vol.~34, no.~3, pp. 173--204, 2022.

\bibitem{knoca2025}
KNOCA, ``French citizens' convention on the climate (la convention citoyenne pour le climat),'' \url{https://www.knoca.eu/national-assemblies/french-citizens-convention-on-the-climate}, n.d.

\bibitem{zanatta2024}
R.~A.~F. Zanatta and M.~Rielli, ``The artificial intelligence legislation in brazil: Technical analysis of the text to be voted on in the federal senate plenary,'' 2024.

\bibitem{digitalpolicy2024}
D.~P. Alert, ``Opened consultation on anpd artificial intelligence and data protection regulatory project,'' \url{https://digitalpolicyalert.org/event/24272-opened-consultation-on-anpd-artificial-intelligence-and-data-protection-regulatory-project}, 2024.

\bibitem{dataprivacy2025}
D.~P. Brasil, ``The artificial intelligence legislation in brazil: Technical analysis of the text to be voted on in the federal senate plenary,'' \url{https://www.dataprivacybr.org/en/the-artificial-intelligence-legislation-in-brazil-technical-analysis-of-the-text-to-be-voted-on-in-the-federal-senate-plenary/}, n.d.

\bibitem{eu2024}
E.~Commission, ``First draft general-purpose ai code of practice,'' \url{https://ec.europa.eu/futurium/en/ai-general-purpose}, 2024.

\bibitem{cowan1997}
R.~S. Cowan, \emph{A Social History of American Technology}.\hskip 1em plus 0.5em minus 0.4em\relax Oxford University Press, 1997.

\end{thebibliography}
\end{document}